\documentclass[%
 aip,
 amsmath,amssymb,
 reprint,%
]{revtex4-1}

\usepackage{braket}
\usepackage{graphicx}
\usepackage{dcolumn}
\usepackage{bm}

\usepackage[utf8]{inputenc}
\usepackage[T1]{fontenc}
\usepackage{mathptmx}
\usepackage{etoolbox}

\makeatletter
\def\@email#1#2{%
 \endgroup
 \patchcmd{\titleblock@produce}
  {\frontmatter@RRAPformat}
  {\frontmatter@RRAPformat{\produce@RRAP{*#1\href{mailto:#2}{#2}}}\frontmatter@RRAPformat}
  {}{}
}%
\makeatother
\begin{document}

\preprint{AIP/123-QED}

\title{Nonadiabatic dynamics near metal surfaces under Floquet engineering: Floquet electronic friction vs. Floquet surface hopping}
\author{Yu Wang}
\email{wangyu19@westlake.edu.cn}
 \affiliation{Department of Chemistry, School of Science, Westlake University, Hangzhou 310024 Zhejiang, China}
 \affiliation{Institute of Natural Sciences, Westlake Institute for Advanced Study, Hangzhou 310024 Zhejiang, China}
\author{Wenjie Dou}%
 \email{douwenjie@westlake.edu.cn}
 \affiliation{Department of Chemistry, School of Science, Westlake University, Hangzhou 310024 Zhejiang, China}
 \affiliation{Institute of Natural Sciences, Westlake Institute for Advanced Study, Hangzhou 310024 Zhejiang, China}
 \affiliation{Department of Physics, School of Science, Westlake University, Hangzhou 310024 Zhejiang, China}

\date{\today}

\begin{abstract}
In the previous study (arXiv:2303.00479), we have derived a Floquet classical master equation (FCME) to treat nonadiabatic dynamics near metal surfaces under Floquet engineering. We have also proposed trajectory surface hopping algorithm to solve the FCME. In this study, we map the FCME into a Floquet Fokker-Planck equation in the limit of fast Floquet driving and fast electron motion as compared to nuclear motion. The Fokker-Planck equation is then being solved using Langevin dynamics with explicit friction and random force from the nonadiabatic effects of hybridized electron and Floquet states. We benchmark the Floquet electronic friction (FEF) dynamics against Floquet quantum master equation (FQME) and Floquet surface hopping (FSH). We find that Floquet driving results in violation of the second fluctuation-dissipation theorem, which further gives rise to heating effects.

\end{abstract}

\maketitle

\section{\label{sec:level1}Introduction}
Floquet engineering is referred to as controlling the quantum systems with time-periodic external fields,
which can give rise to various phenomena in limit of the strong field regime\cite{oka2019floquet, sentef2020quantum}.
To model the system under Floquet engineering, we employ periodic Hamiltonian, $H(t+T)=H(t)$. Here, $T$ is the periodicity ($T=2\pi/\Omega$) and $\Omega$ is the driving frequency\cite{engelhardt2021dynamical}.
Recently, Floquet engineering is being realized by strong light-matter interactions
\cite{gunter2009sub,engel2012light,ebbesen2016hybrid,forn2019ultrastrong,garcia2021manipulating}.
The strong light-matter interactions can result in the hybrid states, termed as the polaritonic states. 
Recent studies show that polaritons strongly modify photophysical and photochemical
processes, 
including the enhancement of vibrational energy transfer
\cite{coles2014polariton,georgiou2018control,xiang2020intermolecular},
maximizing superconducting current
\cite{gimeno2020enhanced},
reducing energy losses in photovoltaics
\cite{nikolis2019strong,wang2021polariton},
tilting the ground-state reactivity landscape
\cite{thomas2019tilting,galego2019cavity,lather2019cavity}, etc.

The problem we concerned in this study is the Floquet engineered nonadiabatic (electron transfer and vibrational relaxation)
processes at the molecule-metal interface\cite{cabra2020optical}, which attracts 
broad interests in the fields of 
substrate-mediated surface photochemistry (or photocatalysis)
\cite{zada2020surface,fang2021recent},
dye-sensitized photovoltaic
\cite{bisquert2004determination,sharma2018dye}, 
or chemisorptions\cite{hammer1996co,kong2018porphyrin,ren2019k}. In the previous work\cite{wang2023nonadiabatic}, we have derived a Floquet classical master equation (FCME) to describe Floquet engineered nonadiabatic dynamics near metal surfaces. We have also proposed a Floquet surface hopping (FSH) algorithm to solve the FCME, where we evolve trajectories on the potential energy surfaces (PESs) with stochastic hopping between different PESs. The hopping rates are determined the Floquet replicas modified molecule-metal interactions.  
This method is valid when two conditions are met:
1) high temperature limit ($\hbar\omega\ll kT$) so that the nuclear motion can be treated classically;
2) weak molecule-metal coupling ($\Gamma\ll kT$) so that 
the effect of molecular level broadening can be disregarded. We have also benchmarked our FSH algorithm, where FSH agrees with FQME well as long as the nuclei can be treated classically regardless the driving amplitude and driving frequency.  

In this paper, we show that in the limit of strong molecule-metal coupling ($\hbar\omega\ll\Gamma$) and fast Floquet driving ($\hbar\omega\ll\hbar\Omega$),
the FCME can be mapped onto a Floquet Fokker-Planck (FFP) equation
with Floquet replicas modified damping force and random force.
This FFP equation can be then solved easily by Langevin dynamics with explicitly Floquet electronic friction (FEF). It is known that electronic friction theory has been widely implemented in nonadiabatic processes at metal surface\cite{box2020determining,lu2019semi,litman2022dissipative}. Here, 
we benchmark the FEF against FQME and FSH for electronic population and nuclear kinetics dynamics under different Floquet drivings amplitudes ($A$)
and frequencies ($\Omega$). 
We find that FEF fails to capture the oscillation feature caused by Floquet drivings at small driving frequencies. 
By contrast, FEF agrees well with FQME and FSH under fast Floquet drivings regardless of driving amplitudes.
We further observe that the violation of the second fluctuation-dissipation theorem induced by Floquet driving leads to the heating effect of nuclear motion, especially under strong driving amplitude.

The structure of this paper is organized as follows.
In Sec. \uppercase\expandafter{\romannumeral2},
we show the derivation of the FFP equation from the FCME.
In Sec. \uppercase\expandafter{\romannumeral3},
we benchmark the dynamics for electronic population as well as nuclear kinetic energy from the FFP equation against FQME and FSH methods. 
Finally, we conclude in Sec. \uppercase\expandafter{\romannumeral4}

\section{Theory}
\subsection{Floquet Classical master equation (FCME)}
We start from the the Anderson-Holstein (AH) model with 
the periodic drivings act on the impurity energy level (molecule), 
which is coupled both to a vibrational degree of freedom (DoF) and a continuum of electronic states:
\begin{equation}\label{eq0}
    \hat{H} = \hat{H}_S + \hat{H}_B + \hat{H}_T
\end{equation}
\begin{equation}\label{eq1}
    \hat{H}_S = (E(x) + A\sin(\Omega t))d^+d  + V_0(x) + \frac{p^2}{2M}
\end{equation}
\begin{equation}\label{eq2}
    \hat{H}_B = \sum_k \epsilon_k c_k^+ \epsilon_k
\end{equation}
\begin{equation}\label{eq3}
    \hat{H}_T = \sum_k V_k(d^+ c_k + c_k^+ d)
\end{equation}
Here $d (d^+)$ and $c_k (c_k^+)$ are the annihilation (creation) operators for 
an electron in the impurity (subsystem)
and in the continuum (bath),  
$E(x)$ is the on-site energy for the impurity that depends on nuclear position.
$V_0(x)$ is the diabatic potential energy surface (PES) for the unoccupied state. 
we can further define the diabatic PES for the (time-independent) occupied state as
$V_1(x)=V_0(x)+E(x)$. 
The periodic driving acts on the impurity energy level with 
a driving amplitude $A$ and 
a driving frequency $\Omega$. 
Without loss of generality, we assume that $V_0(x)$ is taken the form of Harmonic oscillator: 
\begin{equation}\label{eqV0}
    {V}_0(x) = \frac{1}{2}M\omega^2 x^2
\end{equation}

In the FCME, we define the classical phase space probability densities $P_0(x,p,t) (P_1(x,p,t))$ for the nuclear DoFs with the impurity level being unoccupied (occupied). The time evolution of phase space probability densities is given by
\begin{equation}\label{eqCME_p0}
\begin{split}
    \frac{\partial P_0(x,p,t)}{\partial t} = & \frac{\partial V_0(x)}{\partial x}\frac{\partial P_0(x,p,t)}{\partial p} - \frac{p}{m}\frac{\partial P_0(x,p,t)}{\partial x} \\ & - \gamma_{0\rightarrow1}(t)P_0(x,p,t) + \gamma_{1\rightarrow0}(t)P_1(x,p,t)
\end{split}
\end{equation}
\begin{equation}\label{eqCME_p1}
\begin{split}
    \frac{\partial P_1(x,p,t)}{\partial t} = & \frac{\partial V_1(x)}{\partial x}\frac{\partial P_1(x,p,t)}{\partial p} - \frac{p}{m}\frac{\partial P_1(x,p,t)}{\partial x} \\ & + \gamma_{0\rightarrow1}(t)P_0(x,p,t) - \gamma_{1\rightarrow0}(t)P_1(x,p,t)
\end{split}
\end{equation}
where
\begin{equation}\label{eqrate01}
\begin{split}
    \gamma_{0\rightarrow1}(t) = \frac{\Gamma}{\hbar}\tilde{f}(E(x))
\end{split}
\end{equation}
\begin{equation}\label{eqrate10}
\begin{split}
    \gamma_{1\rightarrow0}(t) = \frac{\Gamma}{\hbar}(1-\tilde{f}(E(x)))
\end{split}
\end{equation}
Here $\tilde{f}(E(x))$ is the modified Fermi function with Floquet replicas, which is given by
\begin{equation}\label{eqfermi_floquet}
\begin{split}
    \tilde{f}(E(x)) = & \sum_{nm}\cos\{(n-m)(\Omega t+\pi/2)\} \times \\ &
    J_n(z)J_m(z)\frac{1}{1+e^{\beta(E(x)-m\Omega)}}
\end{split}
\end{equation}
where $n,m$ are integers ranging from $-\infty$ to $+\infty$. $J_n(z)$ is the $n$-th Bessel function of the first kind with $z=\frac{A}{\hbar\Omega}$.

In the limit of fast driving, we can perform the time average on $\tilde{f}(E(x))$, such that we arrive at a time-independent $\bar{\tilde{f}}(E(x))$ 
\begin{equation}
\begin{split}
    \bar{\tilde{f}}(E(x)) = \sum_{n}|J_n(z)|^2\frac{1}{1+e^{\beta(E(x)-n\Omega)}}
\end{split}
\end{equation}
Correspondingly, time-averaged hopping rate $\bar{\gamma}_{0\rightarrow1}(\bar{\gamma}_{1\rightarrow0})$ is also time-independent.
Here, $\Gamma$ is the hybridization function given by
 \begin{equation}\label{eq46}
\begin{split}
    \Gamma(\epsilon) = 2\pi\sum_k|V_k|^2\delta(\epsilon_k-\epsilon)
\end{split}
\end{equation}
In the wide band limit, we can assume that $\Gamma$ is a constant (i.e., does not change with $\epsilon$ or x).

The equation of motion for the phase space densities in this FCME 
can be solved via a Floquet surface hopping algorithm in real time, which is denoted as Floquet averaged surface hopping with density (FaSH-density) algorithm. In short, in the FaSH-density algorithm, we use time-averaged $\bar{\gamma}_{0\rightarrow1}(\bar{\gamma}_{1\rightarrow0})$ as the hopping rates to propagate nuclear dynamics, whereas we use time-dependent hopping rate to propagate electronic dynamics via $\dot{P_0}= - \gamma_{0\rightarrow1}P_0 + \gamma_{1\rightarrow0}P_1$ and 
$\dot{P_1}= \gamma_{0\rightarrow1}P_0 - \gamma_{1\rightarrow0}P_1$. The electronic population is then calculated using $P_0$ and $P_1$. See Ref. \cite{wang2023nonadiabatic} for details.

\subsection{Floquet Fokker-Planck equation (FFP)}
We now map the Floquet CME into a Floquet FP equation with explicit electronic friction and random force. To do so, we first  define new densities $A(x, p, t)$ and $B(x, p, t)$ as 
\begin{equation}\label{eqp0AB}
\begin{split}
    P_0(x, p, t) = (1-\bar{\tilde{f}}(E(x)))A(x, p, t)+ B(x, p, t)
\end{split}
\end{equation}
\begin{equation}\label{eqp1AB}
\begin{split}
    P_1(x, p, t) = \bar{\tilde{f}} (E(x)) A(x, p, t)- B(x, p, t)
\end{split}
\end{equation}
Note that $A(x, p, t)=P_0(x, p, t)+P_1(x, p, t)$, which is the total probability density. When the electronic motion and Floquet driving are very fast, the phase space densities $P_0$ and $P_1$ will be very close to equilibrium densities $(1-\bar{\tilde{f}} (E(x))) A$ and $\bar{\tilde{f}}(E(x))A$ respectively. Such that $B(x, p, t)$ can be seen as the nonadiabatic phase space density. 

The time evolution of $A(x, p, t)$ can be obtained by plugging Eqs. (\ref{eqp0AB}) and (\ref{eqp1AB}) into Eqs. (\ref{eqCME_p0}) and (\ref{eqCME_p1}),
 \begin{equation}\label{eqA}
\begin{split}
    &\frac{\partial A(x,p,t)}{\partial t} \\ = & -\frac{p}{M} \frac{\partial A(x,p,t)}{\partial x}
    \\& + (\frac{\partial V_0(x)}{\partial x} + \frac{d (E(x))}{d x}\bar{\tilde{f}})\frac{\partial A(x,p,t)}{\partial p}
    \\& - \frac{d (E(x))}{d x}\frac{\partial B(x,p,t)}{\partial p}
\end{split}
\end{equation}
The time evolution of $B(x, p, t) = \bar{\tilde{f}}(E(x))P_0(x, p, t) - (1-\bar{\tilde{f}}(E(x)))P_1(x, p, t)$ can be formulated as,
 \begin{equation}\label{eqB}
\begin{split}
    &\frac{\partial B(x,p,t)}{\partial t} \\= & - \frac{p}{M} \frac{\partial B(x,p,t)}{\partial x} + \frac{\partial V_0(x)}{\partial x} \frac{\partial B(x,p,t)}{\partial p}  \\ & +\frac{p}{M} A(x,p,t)\frac{\partial \bar{\tilde{f}} }{\partial x} 
    \\& - \frac{d (E(x))}{d x} \bar{\tilde{f}} ( 1- \bar{\tilde{f}})\frac{\partial A(x,p,t)}{\partial p}
    \\ & + \frac{d (E(x))}{d x}  ( 1- \bar{\tilde{f}}) \frac{\partial B(x,p,t)}{\partial p} \\ & - \Gamma B(x,p,t) - \Gamma ( \tilde f - \bar{\tilde{f}}) A 
\end{split}
\end{equation}
Note that we only invoke the fast driving approximation ($\Omega>\omega$) so far, such that Eqs. (\ref{eqA}) and (\ref{eqB}) are exact as long as FCME is valid. We now invoke the assumption of fast electronic motion as compared to nuclear motion ($\Gamma > \hbar\omega$). In such a limit, the phase space densities are very close to equilibrium densities. Hence, $B(x,p,t)$ should be small relative to $A(x,p,t)$ and change slowly with respect to $x, p, t$. Such that several terms in Eq. (\ref{eqB}) can be ignored,
 \begin{equation}\label{eqBsimple}
\begin{split}
    B(x,p,t) \approx & - \frac{d (E(x))}{d x} \frac{1}{\Gamma}  \bar{\tilde{f}} ( 1- \bar{\tilde{f}})\frac{\partial A(x,p,t)}{\partial p}
    \\ & + \frac{p}{\Gamma M} A(x,p,t)\frac{\partial \bar{\tilde{f}} }{\partial x} - (\tilde f -\bar{\tilde{f}}) A
\end{split}
\end{equation}
If we substitute Eq. (\ref{eqBsimple}) back into Eq. (\ref{eqA}), we arrive at a FFP equation with periodic driving system,
\begin{equation}\label{eqFP0}
\begin{split}
    \frac{\partial A(x,p,t)}{\partial t} = & -\frac{p}{M} \frac{\partial A(x,p,t)}{\partial x}
     +  \frac{\partial U(x,t)}{\partial x}\frac{\partial A(x,p,t)}{\partial p}
    \\& + \gamma_e(x) \frac{\partial}{\partial p}(pA(x,p,t))
    \\& + D(x) \frac{\partial^2A(x,p,t)}{\partial p^2}
\end{split}
\end{equation}
Here, $\gamma_e(x)$ is the electronic friction coefficient,
\begin{equation}\label{eqgamma_e}
\begin{split}
   \gamma_e(x) & =  -\frac{1}{\Gamma M}\frac{dE(x)}{dx}
   \frac{\partial \bar{\tilde{f}} }{\partial x}
\end{split}
\end{equation}
the correlation function of the random force is $D(x)=\gamma_e'(x) MkT$. We have defined $\gamma_e'(x)$ as,
\begin{equation}\label{eqgamma_e'}
\begin{split}
   \gamma_e'(x) = \frac{\beta}{\Gamma M}\left( \frac{dE(x)}{dx} \right)^2   \bar{\tilde{f}} (1-\bar{\tilde{f}} )
\end{split}
\end{equation}
and $\frac{\partial U(x,t)}{\partial x}$ is the time-dependent mean force
\begin{equation}\label{eqdU_real}
\begin{split}
   \frac{\partial U(x,t)}{\partial x}  =  \hbar\omega x + \frac{d (E(x))}{dx}\tilde{f}
\end{split}
\end{equation}
We can write the time-dependent potential of mean force (PMF) $U(x,t)$ explicitly as (up to a constant),
\begin{equation}\label{eqU}
\begin{split}
   U(x,t) = & \frac{1}{2}\hbar\omega x^2 - \frac{1}{\beta}\sum_{nm}\cos\{(n-m)(\Omega t+\pi/2)\} \times \\ & J_n(z)J_m(z) \log(1+\exp(-\beta(E(x)-m\Omega))
\end{split}
\end{equation}
If we further invoke time average on the PMF, we arrive at the time-averaged PMF (aPMF) as
\begin{equation}\label{eq_a_U}
\begin{split}
   \bar{U}(x) = \frac{1}{2}\hbar\omega x^2 - \frac{1}{\beta}\sum_{n}|J_n(z)|^2 \log(1+\exp(-\beta(E(x)-n\Omega))
\end{split}
\end{equation}

This FFP can be solved via
a Floquet electronic friction-Langevin dynamics either with time-dependent potential energy surface $U(x,t)$ (FEF-force), or with time-independent $\bar{U}(x)$ (FEF)
\begin{equation}\label{eqLD1}
\begin{split}
   \dot{p} = - \frac{\partial U}{\partial x} - \gamma_ep + \xi ,
\end{split}
\end{equation}
\begin{equation}\label{eqLD2}
\begin{split}
   \dot{x} =  \frac{p}{M}
\end{split}
\end{equation}
where $\xi$ is the random force that is assumed to be a
Gaussian variable with a norm $\sigma=\sqrt{2M\gamma_e'kT/dt}$.
Again, $dt$ is the time step interval.
We use 4th Runge-Kutta to integrate Eqs. (\ref{eqLD1}) and (\ref{eqLD2}),
and 10000 trajectories are used for both FSH and FEF simulations.

\begin{figure*}[htbp]
\centering
\includegraphics[scale=0.35]{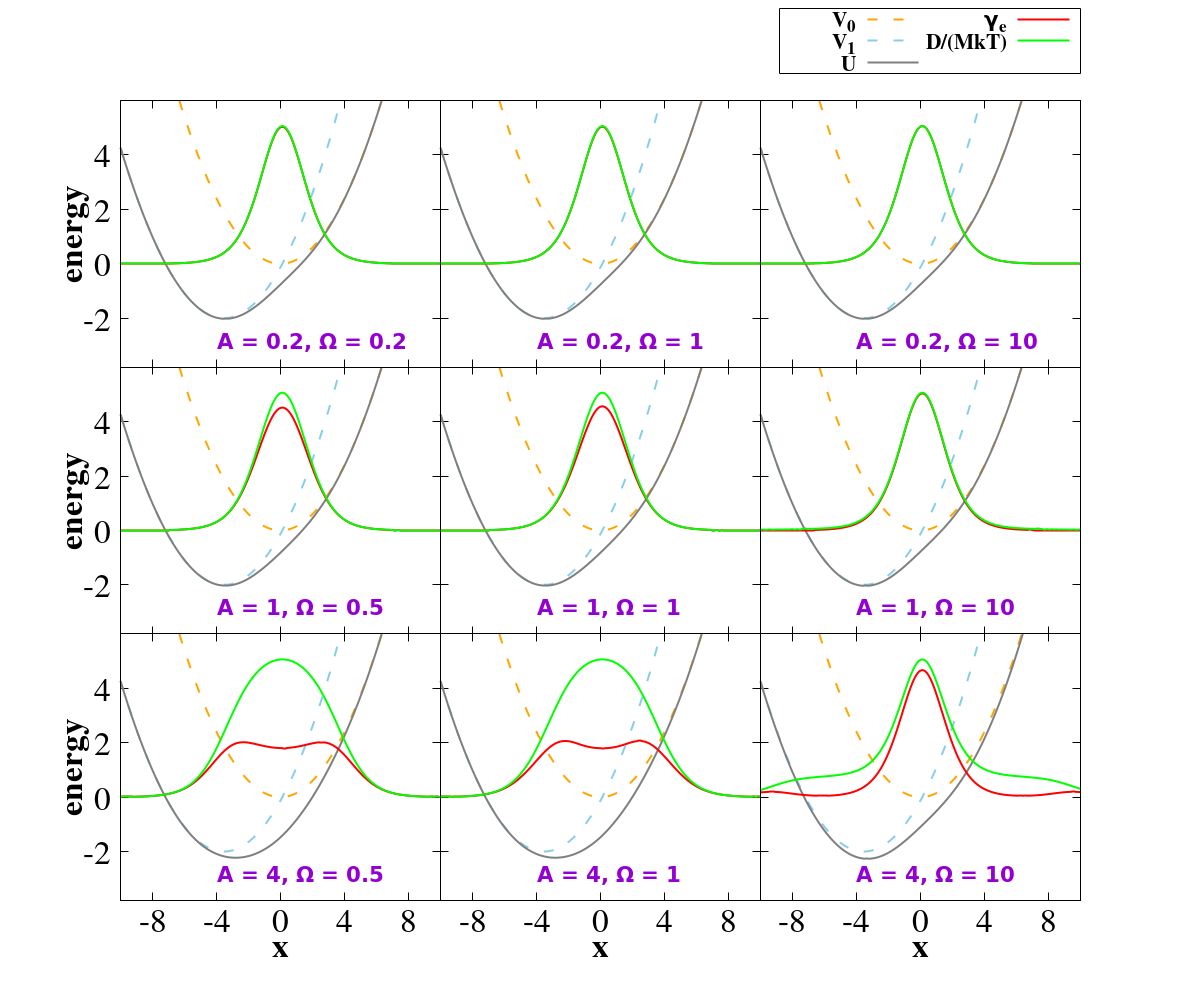}
\caption{Diabatic potential $V_0$ (Eq. \ref{eqV0}, yellow dash line) and $V_1$ (blue dash line), time-averaged potential of mean force (aPMF) $\bar{U}(x)$ (Eq. \ref{eq_a_U}, grey solid line), electronic friction $\gamma_e(x)$ (Eq. \ref{eqgamma_e}, red solid line), as well as $\gamma_e'(x)$ (Eq. \ref{eqgamma_e'}, green solid line) as a function of position $x$ under different driving amplitudes $A$ and driving frequencies $\Omega$. $g=0.75$, $\hbar\omega=0.3$, $\Gamma=1$, $\overline{E}_d=-2$, $kT=1$}
\label{fig_pot}
\end{figure*}

Eqs. (\ref{eqFP0})-(\ref{eq_a_U}) are the main results of this paper.
To better understand these formulas, $E(x)$ is chosen to be linear dependence on $x$,
\begin{equation}
    E(x) = \sqrt{2}gx + E_d
\end{equation}
We define the renormalized energy as $\bar{E}_d\equiv E_d-E_r$, where $E_r=g^2/\hbar\omega$ is the reorganization energy. In Figure \ref{fig_pot}, we plot the aPMF $\bar{U}(x)$ (Eq. \ref{eq_a_U}), electronic friction $\gamma_e(x)$ (Eq. \ref{eqgamma_e}), as well as $\gamma_e'(x)$ (Eq. \ref{eqgamma_e'}) as a function of $x$ under different driving conditions. It is noteworthy that when driving amplitude ($A=1$ or $4$) is much larger than nuclear oscillation ($\hbar\omega=0.3$), $\gamma_e\ne\gamma'_e=\frac{D(x)}{MkT}$, which violates the second fluctuation-dissipation theorem. The heating effects on nuclear motion arise from such violation as seen in the Results section.

\section{Results}
\begin{figure*}[htbp]
\centering
\includegraphics[scale=0.35]{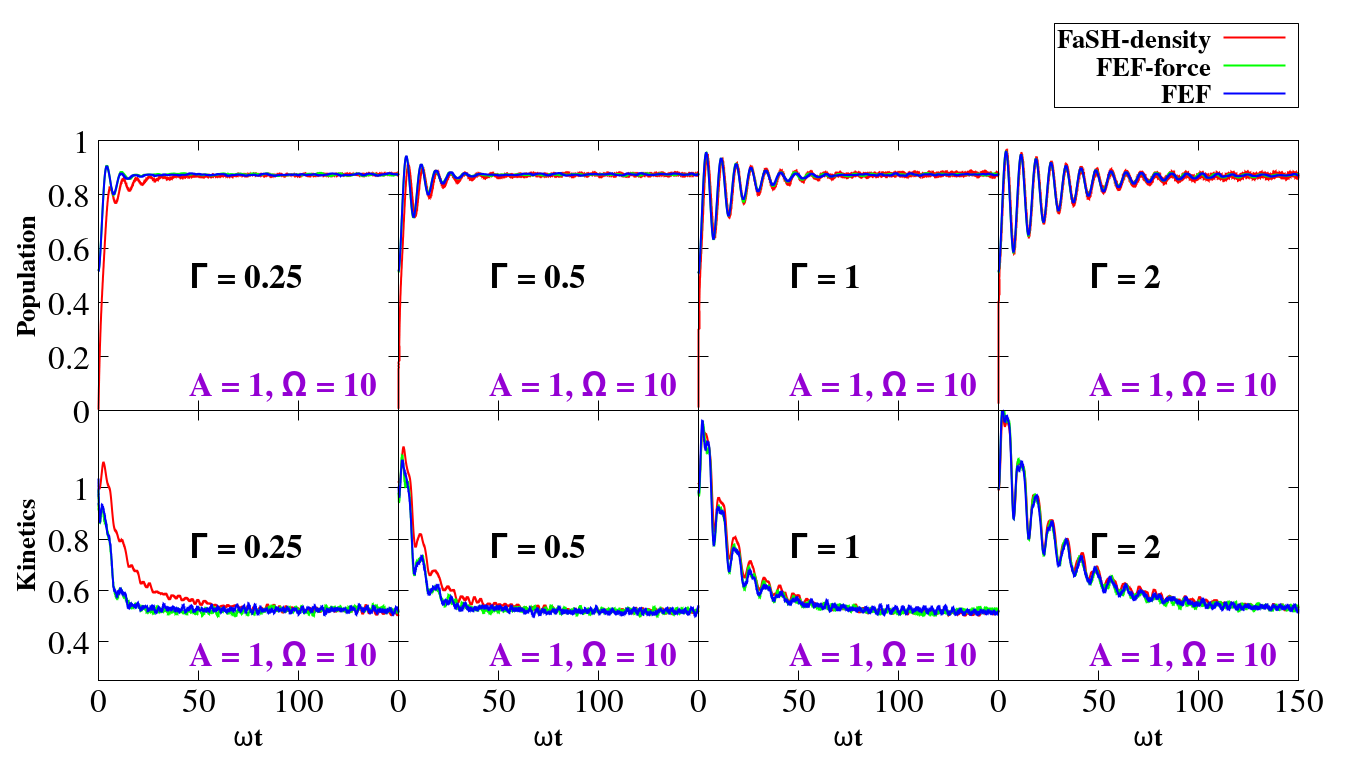}
\caption{Electronic population and nuclear kinetics as a function of time for different $\Gamma$. The driving amplitude and frequency are fixed $A=1, \Omega=10$. We prepare initial state satisfying Boltzmann distribution at temperature $2kT$. Note that FEF(FEF-force) agrees well with FaSH-density in the limit of $\Gamma\gg\hbar\omega$. Parameters: $g=0.75$, $\hbar\omega=0.3$, $\Gamma=1$, $\overline{E}_d=-2$, $kT=1$. }
\label{fig_diff_Gamma}
\end{figure*}

Firstly, we compare the electronic and nuclear dynamics between Floquet surface hopping (FaSH-density) and Floquet electronic friction methods. In Floquet electronic friction, we could either use the   time-dependent PMF (denoted as FEF-force) or the time-independent aPMF (denoted as FEF). In Figure \ref{fig_diff_Gamma}, we benchmark these Floquet electronic friction methods for different $\Gamma$.  
We fix the driving amplitude $A=1$ and driving frequency $\Omega=10$ and prepare the initial states of the oscillators in one well
with a Boltzmann distribution at temperature $2kT$.
Similar to the non-Floquet case \cite{dou2015frictional}, the potential of mean force
is a mixture of two diabatic potential of energy surfaces (PESs), such that 
the initial ($t=0$) electronic populations
from FEF are not equal to 1. In the long time dynamics for the electronic population, Floquet electronic friction agrees with Floquet surface hopping very well. As for the nuclear dynamics, when electronic motion is fast (large $\Gamma$), we reach to good agreements between Floquet surface hopping and Floquet electronic friction. As expected, Floquet electronic friction fails in the slow electronic motion limit (small $\Gamma$). Below, we mainly force on the large $\Gamma$ limit.

We now benchmark the FEF and FaSH-density against Floquet quantum master equation (FQME). The FQME is exact as long as the broadening effects from the lead can be ignored $\Gamma < kT $.  Details on FQME can be found in Ref. \cite{wang2023nonadiabatic}.

\begin{figure*}[htbp]
\centering
\includegraphics[scale=0.35]{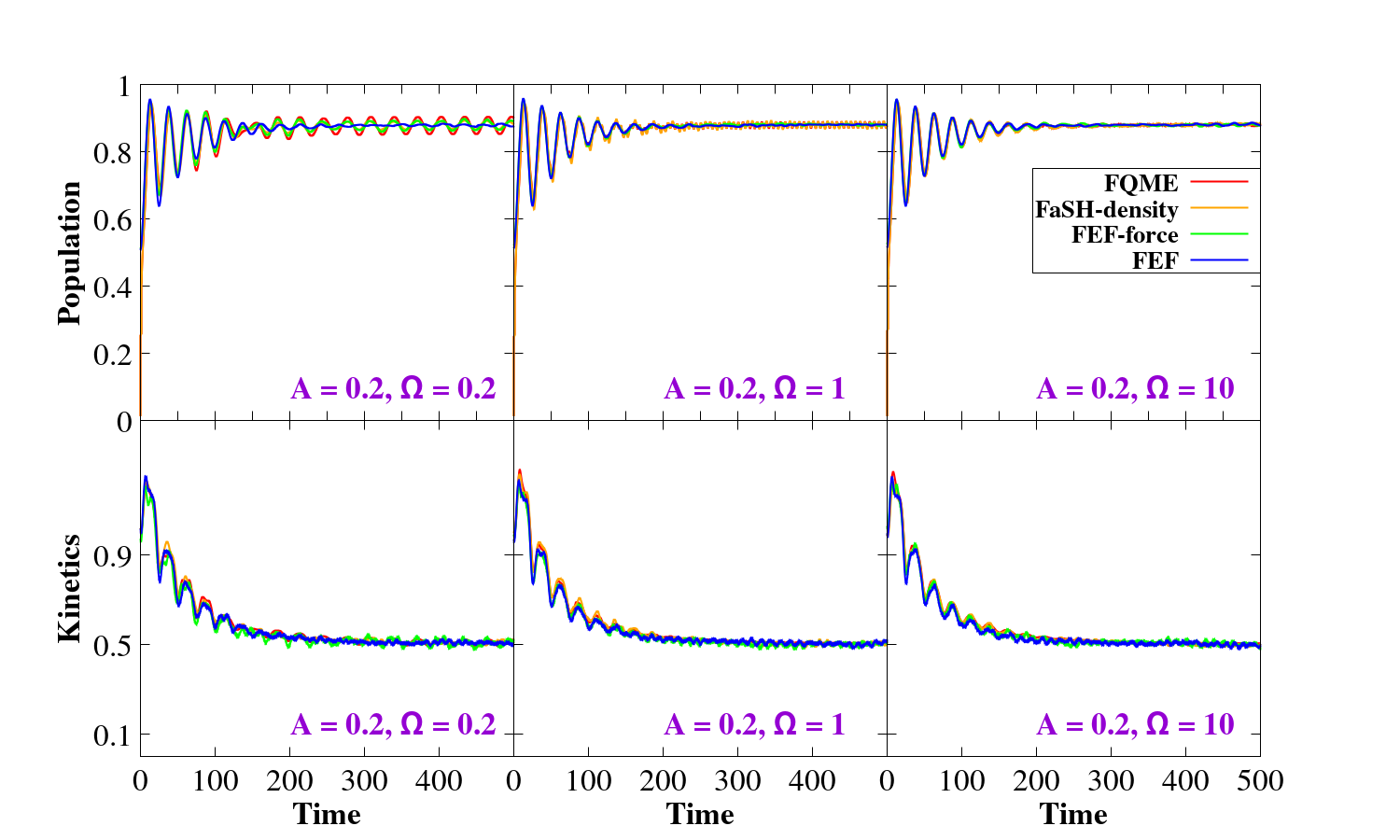}
\caption{Electronic population and nuclear kinetics as a function of time for different driving frequencies ($\Omega=0.2,1,10$) under a small driving amplitude $A=0.2$. $g=0.75$, $\hbar\omega=0.3$, $\Gamma=1$, $\overline{E}_d=-2$, $kT=1$. Note that the four methods (FQME, FaSH-density, FEF-force, and FEF) give nearly the same features of dynamics under small strength of drivings.}
\label{fig_a02}
\end{figure*}

In Figure \ref{fig_a02}, we benchmark the Floquet electronic friction method under relatively small driving amplitude ($A=0.2$), which is comparable to the nuclear oscillation ($\hbar\omega=0.3$). In such a case, all these four methods nearly agree with each other regardless of the driving frequencies. Note that under small driving frequency ($\Omega=0.2$), electronic dynamics reach to a limit cycle, instead of a steady state, which can be reflected by FQME, FaSH-density, as well as FEF-force. FEF method fails to reproduce this limit cycle feature since we average the time-dependent PMF. Under large driving frequency ($\Omega=10$), the electronic populations reach to a steady state of $N=\bar{\tilde{f}}(\bar{E}_d)$, and nuclear kinetics reach to a steady state of $1/2kT$, where $kT$ is the temperature of the metal bath.

\begin{figure*}[htbp]
\centering
\includegraphics[scale=0.35]{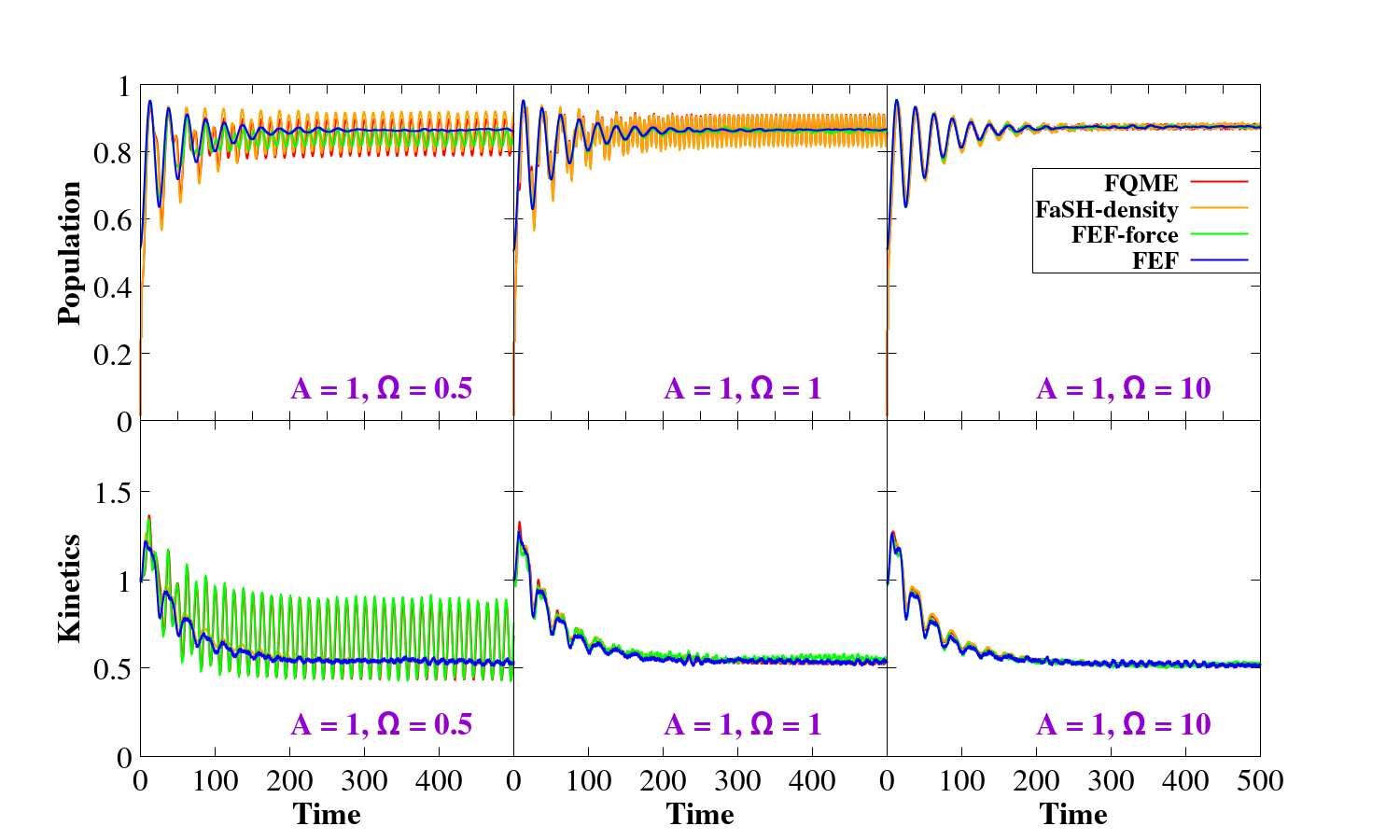}
\caption{Electronic population and nuclear kinetics as a function of time for different driving frequencies ($\Omega=0.5,1,10$) under a medium strength of driving amplitude $A=1$. $g=0.75$, $\hbar\omega=0.3$, $\Gamma=1$, $\overline{E}_d=-2$, $kT=1$. Note that FEF method using time-independent aPMF fails to capture the oscillation feature introduced by the drivings at small driving frequencies ($\Omega=0.5$ or $1$).}
\label{fig_a1}
\end{figure*}

We now turn to the case of medium strength in driving amplitude ($A=1$), as shown in Figure \ref{fig_a1}. The oscillation feature from FQME, FaSH-density, and FEF-force under small driving frequencies ($\Omega=0.5,1$) is more pronounced than the small driving amplitude case (Figure \ref{fig_a02}), which reach into a cycle limit in the long time. Note that FEF-force gives smaller oscillation feature for electronic population in the limit cycle as compared to FQME, whereas FEF method fails to capture these oscillation. In the limit of high driving frequency ($\Omega=10$), all these methods reach to the same steady state for both electronic and nuclear dynamics. 

\begin{figure*}[htbp]
\centering
\includegraphics[scale=0.35]{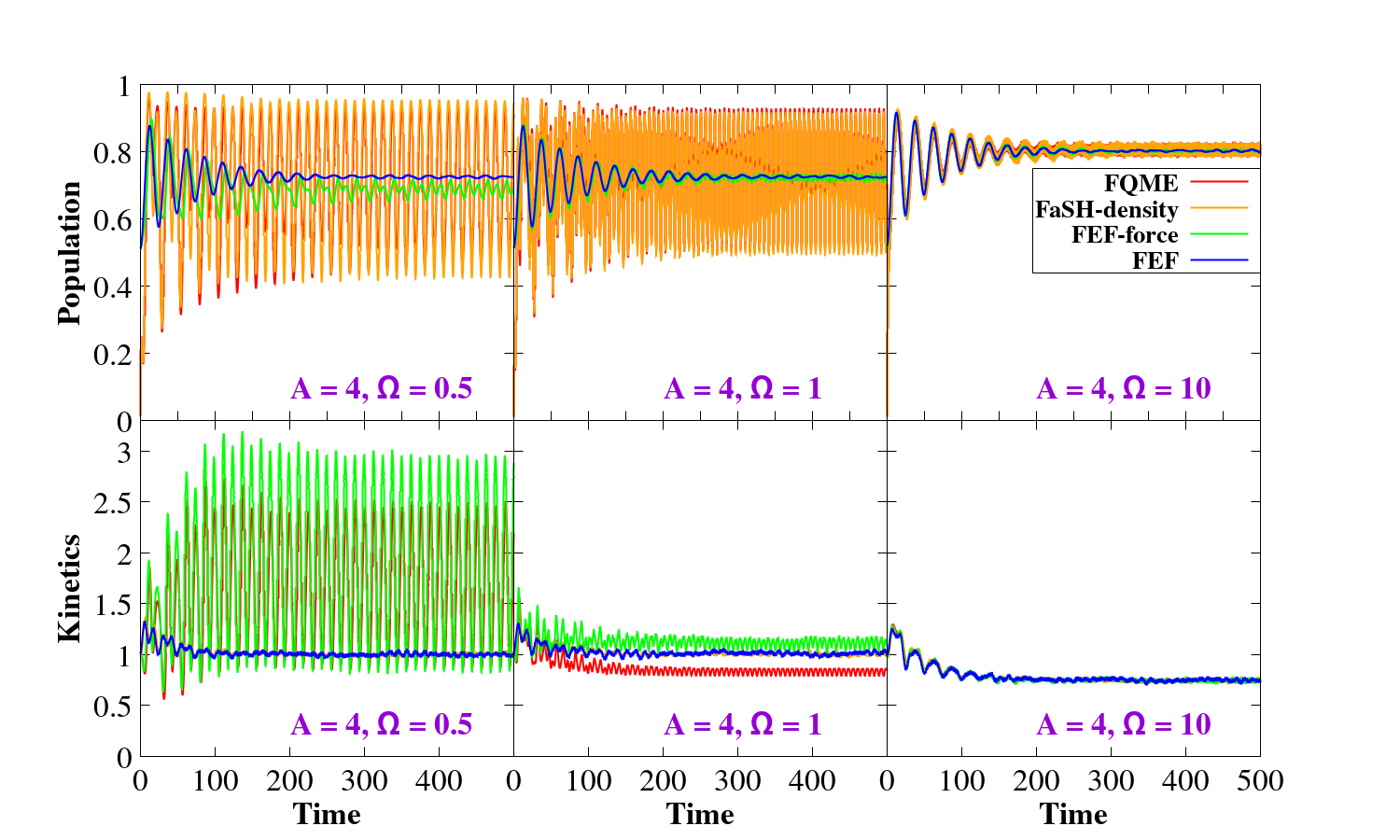}
\caption{Electronic population and nuclear kinetics as a function of time for different driving frequencies ($\Omega=0.5,1,10$) under a strong driving amplitude $A=4$. $g=0.75$, $\hbar\omega=0.3$, $\Gamma=1$, $\overline{E}_d=-2$, $kT=1$. FEF fails to capture the oscillation features when the driving is slow ($\Omega=0.5$ or $1$). FEF-force predicts the oscillator features with smaller oscillating amplitude. The effective temperature of nuclear motion is above $1/2kT$ which arising from the violation of the second fluctuation-dissipation theorem.}
\label{fig_a4}
\end{figure*}

Finally, we show the case where the driving amplitude is relatively large ($A=4$) in Figure \ref{fig_a4}. In this limit, the oscillation feature is very strong, especially when driving frequency is small. Again, FEF fails to capture these oscillation features when the driving is slow. FEF-force predicts the oscillator features with smaller oscillating amplitude. It is noteworthy that under such strong driving amplitude, the nuclear kinetics reach to a steady state above $1/2kT$, which demonstrate the violation of the second fluctuation-dissipation theorem as illustrated in Figure \ref{fig_pot}, thus giving rise to the heating effect of nuclear motion.

\section{Conclusions}
In this paper, we derived a Floquet Fokker-Planck equation (FFP) to characterize Floquet classical master equation (FCME) in the limit of slow nuclear motion as compared to fast electronic motion as well as fast external driving ($\Gamma\gg\hbar\omega$ and $\hbar\Omega\gg\hbar\omega$). The FFP can be solved using Langevin dynamics with explicit Floquet replicas modified electronic friction and random force. By employing time-dependent and time-averaged mean force, we proposed two FEF methods in this study, which agree well with FQME and FaSH-density under fast Floquet drivings regardless of driving amplitudes. We find that Floquet driving lead to a violation of the second fluctuation-dissipation theorem, especially under large driving amplitude. which giving rise to heating effect on nuclear motion. Our method offers an alternative means to study nonadiabatic dynamics with periodic drivings in open quantum system.

\begin{acknowledgments}
This material is based upon work supported by National Natural Science Foundation of China (NSFC No. 22273075). W.D. acknowledges start-up funding from Westlake University. 
\end{acknowledgments}

\bibliography{reference}
\end{document}